\documentclass[conference]{IEEEtran}
\IEEEoverridecommandlockouts
\usepackage{macro}
\usepackage{cite}
\usepackage{amsmath,amssymb,amsfonts}
\usepackage{bm}
\usepackage{algorithmic}
\usepackage{graphicx}
\usepackage{textcomp}
\usepackage{xcolor}
\usepackage{algorithm}
\usepackage{algorithmic}
\usepackage{booktabs}
\usepackage{multirow}
\usepackage{hyperref}
\usepackage{cleveref}
\def\BibTeX{{\rm B\kern-.05em{\sc i\kern-.025em b}\kern-.08em
    T\kern-.1667em\lower.7ex\hbox{E}\kern-.125emX}}

\begin{document}

\title{Tada-DIP: Input-adaptive Deep Image Prior for One-shot 3D Image Reconstruction 
\thanks{$^*$Equal contribution.
This work was supported in part by
the National Science Foundation (NSF) grants CCF-2212065, ECCS-2436945, and the NSF CAREER Award CCF-2442240. EB is supported by the U.S. Department of Energy Computational Science Graduate Fellowship. Code is available on GitHub: \url{https://github.com/evanbell02/Tada-DIP}
\smallskip
\newline{\scriptsize
    © 2025 IEEE. Personal use of this material is permitted. Permission from IEEE must be obtained for all other uses, in any current or future media, including reprinting/republishing this material for advertising or promotional purposes, creating new collective works, for resale or redistribution to servers or lists, or reuse of any copyrighted component of this work in other works.}}
}

\author{\IEEEauthorblockN{Evan Bell$^{1,*}$\quad\quad Shijun Liang$^{1,*}$\quad\quad Ismail Alkhouri$^{1,2}$\quad\quad Saiprasad Ravishankar$^{1,3}$}
\smallskip
\IEEEauthorblockA{$^1$Dept. of Computational Mathematics, Science \& Engineering, Michigan State University, East Lansing, MI, USA\\
$^2$Dept. of Electrical Engineering and Computer Science, University of Michigan, Ann Arbor, MI, USA\\
$^3$Dept. of Biomedical Engineering, Michigan State University, East Lansing, MI, USA} }

\maketitle

\begin{abstract}
Deep Image Prior (DIP) has recently emerged as a promising one-shot neural-network based image reconstruction method. However, DIP has seen limited application to 3D image reconstruction problems. In this work, we introduce \textit{Tada-DIP}, a highly effective and fully 3D DIP method for solving 3D inverse problems. By combining input-adaptation and denoising regularization, Tada-DIP produces high-quality 3D reconstructions while avoiding the overfitting phenomenon that is common in DIP. Experiments on sparse-view X-ray computed tomography reconstruction validate the effectiveness of the proposed method, demonstrating that Tada-DIP produces much better reconstructions than training-data-free baselines and achieves reconstruction performance on par with a supervised network trained using a large dataset with fully-sampled volumes.
\end{abstract}


\section{Introduction}

Three-dimensional image reconstruction problems arise in a number of important applications, including magnetic resonance imaging (MRI)~\cite{fesslermri} and X-ray computed tomography (CT)~\cite{ctsurvey}. The 3D image reconstruction problem can be mathematically formulated as attempting to recover a 3D (vectorized) image $\x\in\R^p$ as accurately as possible from a set of (possibly noisy) measurements $\y\in\R^q$, modeled as
\begin{equation}
    \y = \A\x + \n, \quad \n\sim\N(\0, \sigma^2\I),
    \label{eq:inverse_problem}
\end{equation}
where the forward operator $\A\in\R^{q\times p}$ captures the physics of the imaging system, and $\n$ is an additive noise vector.

When the measurements $\y$ are appreciably undersampled, as in accelerated MRI or sparse-view CT, the reconstruction problem is ill-posed since $q$ is much smaller than $p$~\cite{saisurvey}. Consequently, exploiting prior information about image structure is necessary to obtain high-quality reconstructions. Traditional approaches rely on hand-crafted priors to solve \eqref{eq:inverse_problem}, such as total variation~\cite{asd-pocs}, or wavelet sparsity~\cite{lustig2007sparse}. However, in recent years, machine learning (ML) approaches have emerged as the new state-of-the-art for solving inverse imaging problems~\cite{hammernik2018learning,jin2017deep,chung2023diffusion, alkhourisitcom}.

While these data-driven approaches have demonstrated impressive performance compared to classical methods, they require large (commonly labeled or fully-sampled) datasets to be effective, which are not always available. One ML method that operates without training data is \textit{Deep Image Prior} (DIP)~\cite{ulyanov2018deep}. DIP is a training-data-free method that leverages an \textit{untrained} neural network as an image prior.

DIP and subsequent variants~\cite{liu2019image,bell2023robust,liang2025analysis,alkhouri2024image} (see \cite{alkhouri2025understanding} for a recent tutorial paper) have demonstrated strong performance for one-shot image reconstruction tasks, yet the application of DIP methods to 3D image reconstruction remains relatively unexplored. A small number of previous studies have applied DIP to 3D reconstruction~\cite{gong2018pet,hashimoto2023fully,mayo2024stodip,hisham2025family}. However, in most of these studies, either the reconstructed volume is relatively small, or the DIP method used is not state-of-the-art (no regularization is used or the regularization is total variation or a similar cost function~\cite{gong2018pet,hashimoto2023fully,mayo2024stodip}).

\begin{figure}[!htbp]
    \centering
    \includegraphics[width=\linewidth]{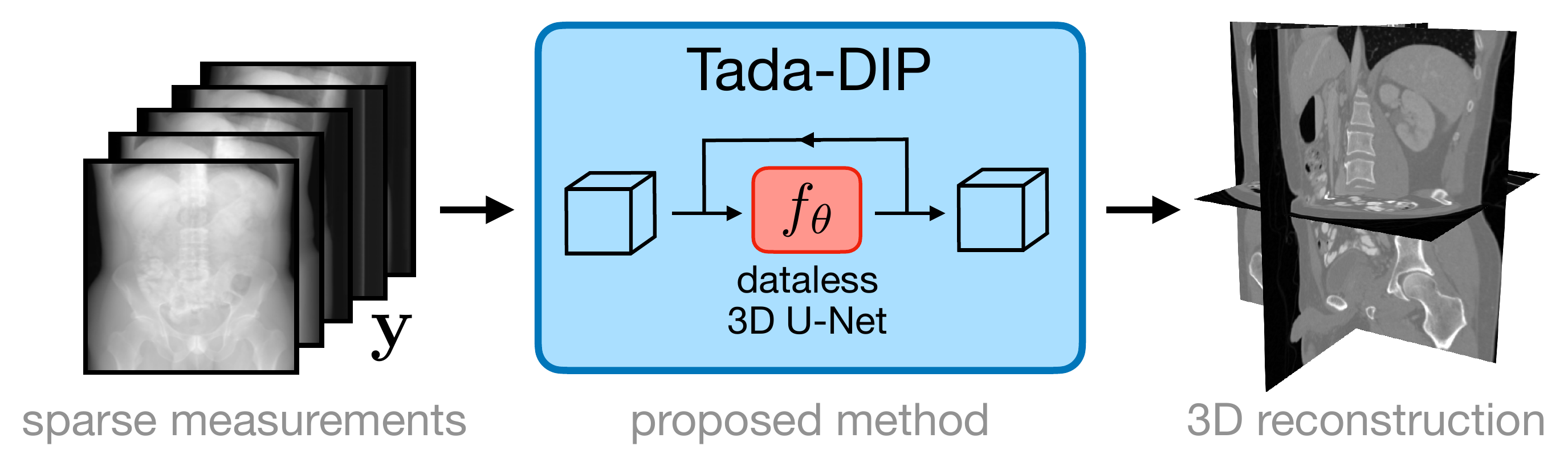}
    \caption{Overview of the setup of the proposed algorithm, \textit{Tada-DIP}.}
    \label{fig:overview}
\end{figure}

\textbf{Contribution.\quad}In this work, we develop a novel DIP-based method for 3D image reconstruction, dubbed Inpu\underline{t-ada}ptive DIP (\textit{Tada-DIP}). The proposed Tada-DIP operates completely in 3D by using a 3D neural network as the backbone (as in previous studies~\cite{gong2018pet,hashimoto2023fully,mayo2024stodip}). The setup of the proposed method is illustrated in \autoref{fig:overview}. By combining input-adaptation and powerful regularization, Tada-DIP produces high-quality 3D reconstructions from single sets of undersampled measurements without relying on external training data. The proposed scheme is introduced in detail in \Cref{sec:method}. We experimentally validate the proposed approach for 3D X-ray CT image reconstruction in \Cref{sec:experiments}. Tada-DIP is used to reconstruct abdominal CT volumes of size $256^3$ from very sparse measurements (30 views and 15 views). To the best of our knowledge, these are the largest images reconstructed with DIP reported in the literature. Moreover, Tada-DIP offers very strong performance. Our results demonstrate that Tada-DIP substantially outperforms dataless baselines for these challenging tasks and even provides reconstructions on par with a \textit{supervised} network trained on a large dataset.

\section{Preliminaries}

\textbf{Deep Image Prior.\quad} Vanilla DIP solves the inverse problem in~\eqref{eq:inverse_problem} by using the following optimization \cite{ulyanov2018deep}:
\begin{equation}
    \hat{\btheta} = \underset{\btheta}{\arg\min}~||\A f_\btheta(\z) - \y||_2^2;\quad \hat{\x} = f_{\hat{\btheta}}(\z),
    \label{eq:dip}
\end{equation}
where $f$ is a neural network (typically a deep U-Net~\cite{unet}) with randomly initialized parameters $\btheta$, and $\z$ is a fixed network input which is typically selected as random noise. While Vanilla DIP can be effective for some problems and settings, certain challenges naturally arise when using DIP. First, it is unclear when to stop the optimization in~\eqref{eq:dip}, since the ground truth image is unknown~\cite{alkhouri2025understanding}. Moreover, running the optimization for too many iterations typically leads to degraded performance, as the network either overfits to noise in the measurements $\y$, or begins to output artifacts in the null space of $\A$~\cite{liang2025analysis}. The second major limitation of Vanilla DIP is that it rarely achieves state-of-the-art performance compared to purpose-built methods for specific imaging tasks. Consequently, developing algorithms that offer improved performance over Vanilla DIP is an important research direction.

\textbf{Prior Art in DIP.\quad}Many previous studies have addressed the problems of preventing overfitting and improving the performance of DIP \cite{alkhouri2025understanding}. There are two key algorithmic components that are present in many of the most successful DIP algorithms. The first key component for a robust and performant DIP algorithm is \textit{regularization}. In particular, \textit{denoising} and \textit{autoencoding} have repeatedly been found to be very effective for DIP. Some DIP algorithms that feature denoising (or noise injection) include DeepRED~\cite{mataev2019deepred}, which uses an external denoiser for regularization, SGLD-DIP~\cite{cheng2019bayesian}, which injects noise into the gradient updates during optimization, and Self-Guided DIP~\cite{liang2025analysis}, which trains $f_\btheta$ to act as a denoiser of its input. Indeed, even the original Deep Image Prior manuscript suggests noise injection as a form of regularization by jittering the network input with additional noise at each iteration~\cite{ulyanov2018deep}. Autoencoding regularization~\cite{alkhouri2024image} has been shown to be similarly effective.

The second key component for a strong DIP algorithm is \textit{updating the network input}. Previous studies have demonstrated both empirically~\cite{zhao2020reference} and theoretically~\cite{alkhouri2024image} that choosing an input $\z$ that is close the to the ground truth $\x$ instead of random noise can boost the performance of DIP. However, in the absence of training data, it is not possible to use a reference image as $\z$. In this case, one can either optimize $\z$ directly to try to minimize the data-fidelity loss, as in Deep Random Projector~\cite{li2023deep} and Self-Guided DIP~\cite{liang2025analysis}, or by setting the network input to the current network output as in aSeq-DIP~\cite{alkhouri2024image} (since the output may be a good estimate of $\x$).

\section{Method: Tada-DIP}
\label{sec:method}

The proposed Tada-DIP incorporates both of the key algorithmic components outlined in the previous section by utilizing denoising regularization and updating the network input at every iteration. Additionally, Tada-DIP is a fully 3D method, so the network $f_\btheta$ is taken to be a 3D U-Net.

To implement the denoising regularization, we inject noise $\boldeta$ into the network input at every iteration. That is, the predicted reconstruction is computed as $\hat{\x} = f_\btheta(\z + \boldeta)$, where $\z$ is the current network input. To encourage $f_\btheta$ to act as a denoiser, the loss function used to update $\btheta$ incorporates a denoising regularization term in addition to the data-fidelity term:
$$
\mathcal{L} = \underbrace{||\y - \A\hat{\x}||_p^p}_{\text{data fidelity}} + \underbrace{\beta\,||\z - \hat{\x}||_p^p}_{\text{denoising reg.}},
$$
where $\beta\in\R$ is a hyperparameter. To draw the noise $\boldeta$, we propose to use Gaussian noise, where the noise level is set according to the scale of $\z$. In particular, at every iteration, the standard deviation of the noise is calculated as $\sigma = \alpha\,\max(|\z|)$, where $|\,\cdot\,|$ is applied element-wise, and $\alpha$ is a hyperparameter. The noise is then drawn as $\boldeta \sim \N(\0, \sigma^2\I)$. This simple scheme prevents the noise level from becoming too large or small relative to the input.

The final component of Tada-DIP is the input update. We propose the simple fixed update rule of setting the updated $\z$ equal to a linear combination of the previous $\z$ and $\hat{\x}$, i.e., $(1-\gamma)\z + \gamma \hat{\x}$, where $\gamma$ is a hyperparameter (we use $\gamma=0.01$). If $\hat{\x}$ is a reasonable estimate of the true $\x$, then we may expect this update to accelerate the optimization and improve the method's performance by injecting some useful information at the network's input.

\begin{figure}[!htbp]
    \centering
    \includegraphics[width=\linewidth]{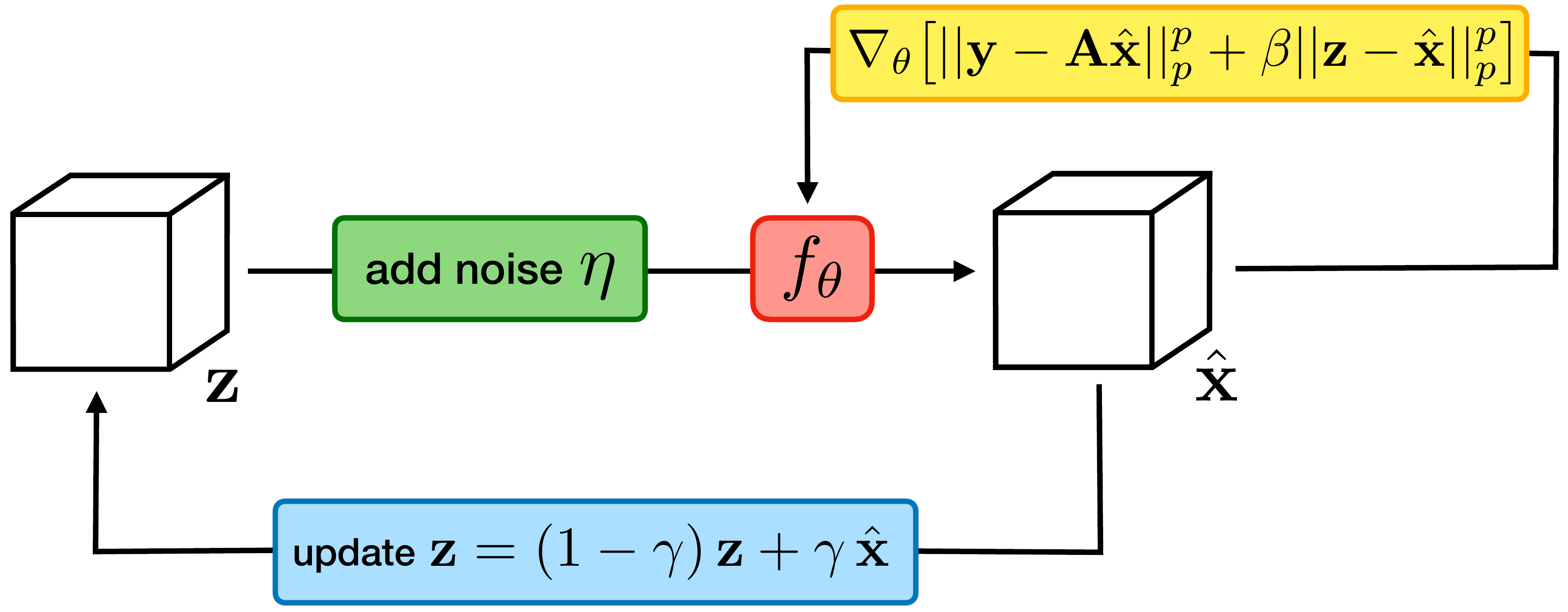}
    \caption{Block diagram illustrating one iteration of the proposed Tada-DIP algorithm.}
    \label{fig:block-diagram}
\end{figure}

The complete Tada-DIP algorithm is given in \Cref{alg:tada_dip}, and one step of the proposed scheme is illustrated in \autoref{fig:block-diagram}. We note that Tada-DIP actually specifies to the previously proposed \textit{aSeq-DIP}~\cite{alkhouri2024image} for a particular choice of hyperparameters ($\alpha=0$, $\gamma=1$), which would turn the regularization term into an autoencoding objective, and update the input by simply setting $\z$ equal to $\hat{\x}$.

\begin{algorithm}
    \small
    \caption{Tada-DIP}\label{alg:tada_dip}
    \begin{algorithmic}
        \STATE \textbf{Input:} measurement $\y=\A\x + \n$, 3D U-Net $f_\btheta$, initial input $\z \sim \N(\0, \I)$, hyperparameters $\alpha$ {\color{gray}$(= 1/2)$}, $\beta$ {\color{gray}$(= 10^{-2})$}, $\gamma$ {\color{gray}$(=10^{-2})$}, $p$ {\color{gray}$(=1)$}, optimizer \texttt{opt} {\color{gray} (Adam)}, optimization iterations $n$ {\color{gray}$(=50 000)$}
        \FOR{iteration = $1$ to $n$}
            \STATE 1. \quad $\sigma \leftarrow \alpha \,\cdot\, \max(|\z|)$
            \STATE 2. \quad $\boldeta \sim \N(\0, \sigma^2\I)$
            \STATE 3. \quad $\hat{\x} \leftarrow f_\btheta(\z + \boldeta)$
            \STATE 4. \quad $\mathcal{L} \leftarrow ||\y - \A\hat{\x}||_p^p + \beta ||\z - \hat{\x}||_p^p$
            \STATE 5. \quad compute $\nabla_\btheta \mathcal{L}$ and update $\btheta$ with \texttt{opt}
            \STATE 6. \quad $\z \leftarrow (1-\gamma)\,\z + \gamma\,\hat{\x}$
        \ENDFOR
        \STATE \textbf{Return:} reconstruction $\hat{\x}$
    \end{algorithmic}
\end{algorithm}

\section{Experiments}
\label{sec:experiments}

\subsection{Experimental Setup}

\textbf{Datasets.\quad}We evaluated Tada-DIP and baseline methods for sparse-view 3D CT reconstruction using a large public dataset provided by the Mayo Clinic, which has been released as the 2021 LDCT Image and Projection dataset~\cite{mayo_dataset} and the 2016 AAPM LDCT Grand Challenge dataset~\cite{aapm_dataset}. We note that we only use the included full dose images. In total, the dataset used contains 209 chest and abdominal CT volumes. The volumes were pre-processed by normalizing each volume so that the voxel values lie in the range $[0, 1]$. Each volume was then resized using trilinear interpolation so that the axial plane had a resolution of $256^2$ (from $512^2$). For evaluation, we used three abdominal CT volumes (cases L067, L096, and L143), which were centrally cropped to a size of $256^3$ after interpolation.

\textbf{Baselines.\quad}We compare Tada-DIP against three training-data-free baselines and one data-driven (supervised) baseline. The unsupervised baselines are filtered backprojection (FBP), total variation (TV) regularized reconstruction (implemented with the adaptive-steepest-descent projection onto convex sets (ASD-POCS)~\cite{asd-pocs} routine in the LEAP library~\cite{leap}), and Vanilla DIP with a 3D U-Net. The supervised baseline is a 2D U-Net which was trained with the FBP reconstruction as input. We note that we also attempted to train a supervised 3D U-Net (with over 160 volumes used for training), but we found that the 2D U-Net performed substantially better (more than 6dB better, in terms of PSNR). We conjecture that the dataset used is not large enough to effectively train a supervised 3D U-Net, even though it is one of the largest publicly available CT datasets.

\textbf{Implementation Details.\quad}All reconstruction methods were tested for 30-view and 15-view CT reconstruction with a parallel beam forward model. For ASD-POCS, we ran the optimization for 500 iterations with 30 subsets and 50 TV steps per iteration. For Vanilla DIP, we used the same 3D U-Net architecture as Tada-DIP, used the $\ell^1$ norm in the data fidelity loss (which performed better than $\ell^2$), and ran the optimization for $50000$ iterations. For Tada-DIP, we set the hyperparameters $\alpha=1/2$, $\beta=0.01$, $\gamma=0.01$, and $p=1$, and  ran Algorithm~\ref{alg:tada_dip} for $50000$ iterations. For both DIP methods, we used the Adam optimizer with a learning rate of $10^{-3}$ and default hyperparameters. We also maintained an exponential moving average of the model outputs for both methods. This moving average was used as the final reconstruction. The supervised U-Net was trained on 80\% of the slices from the processed 2021 LDCT Image data, with 20\% reserved for validation. In total, the training set contained $17091$ slices. 

\subsection{Results}
\begin{figure}[thpb]
    \centering
    \includegraphics[width=\linewidth]{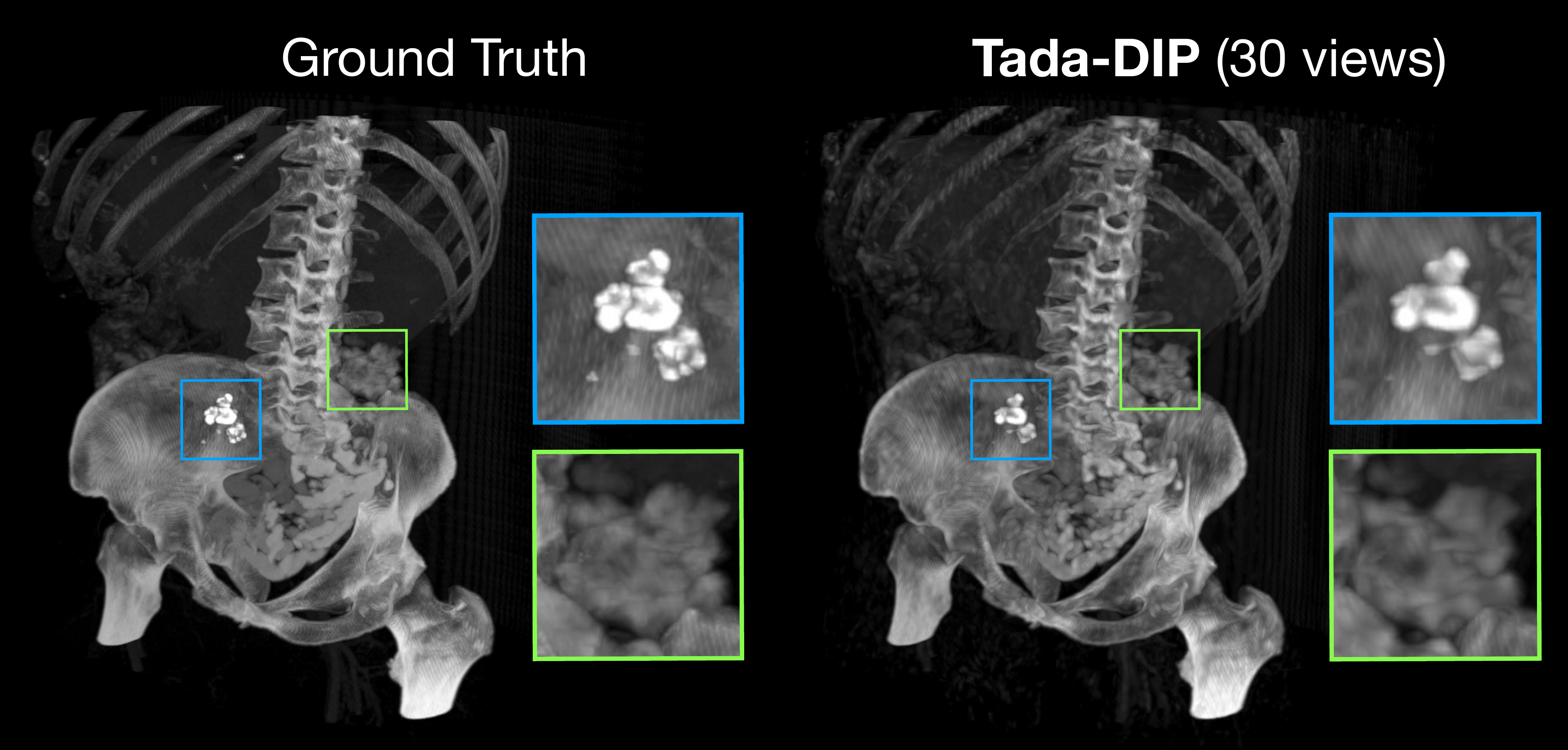}
    \caption{3D visualization (maximum intensity projection) of one $256^3$ test volume reconstructed using the proposed Tada-DIP from 30 views. The softer tissues have been thresholded away. Zoom-ins highlight Tada-DIP's ability to faithfully reconstruct fine details.}
    \label{fig:3d_visualization}
\end{figure}

\begin{figure}[htbp]
    \centering
    \includegraphics[width=\linewidth]{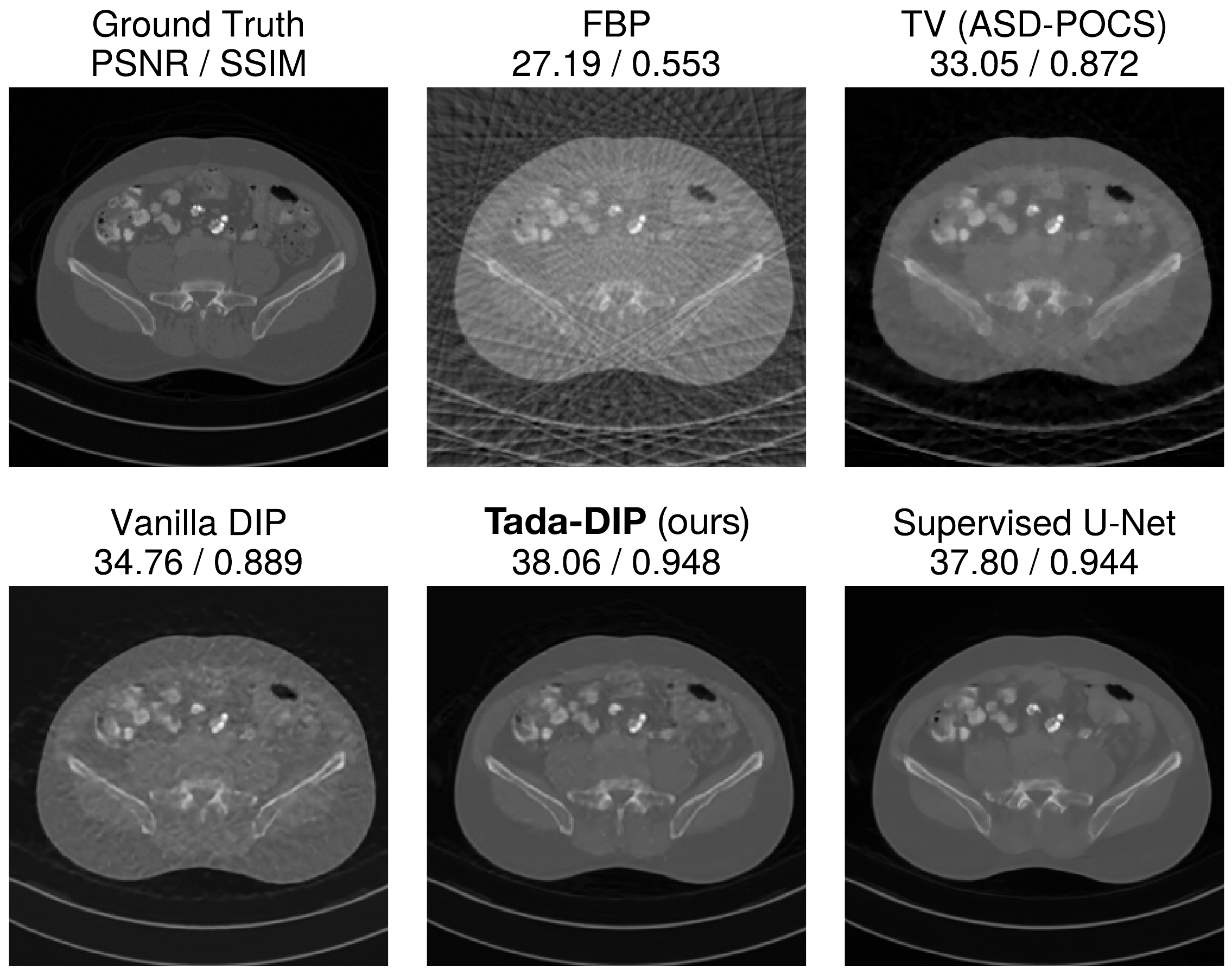}
    \caption{Qualitative comparison of reconstruction methods for 30-view parallel beam CT reconstruction. The visualization shows one axial slice of the 3D reconstruction.}
    \label{fig:qual_comp_30_axial}
\end{figure}

\begin{figure}[htbp]
    \centering
    \includegraphics[width=\linewidth]{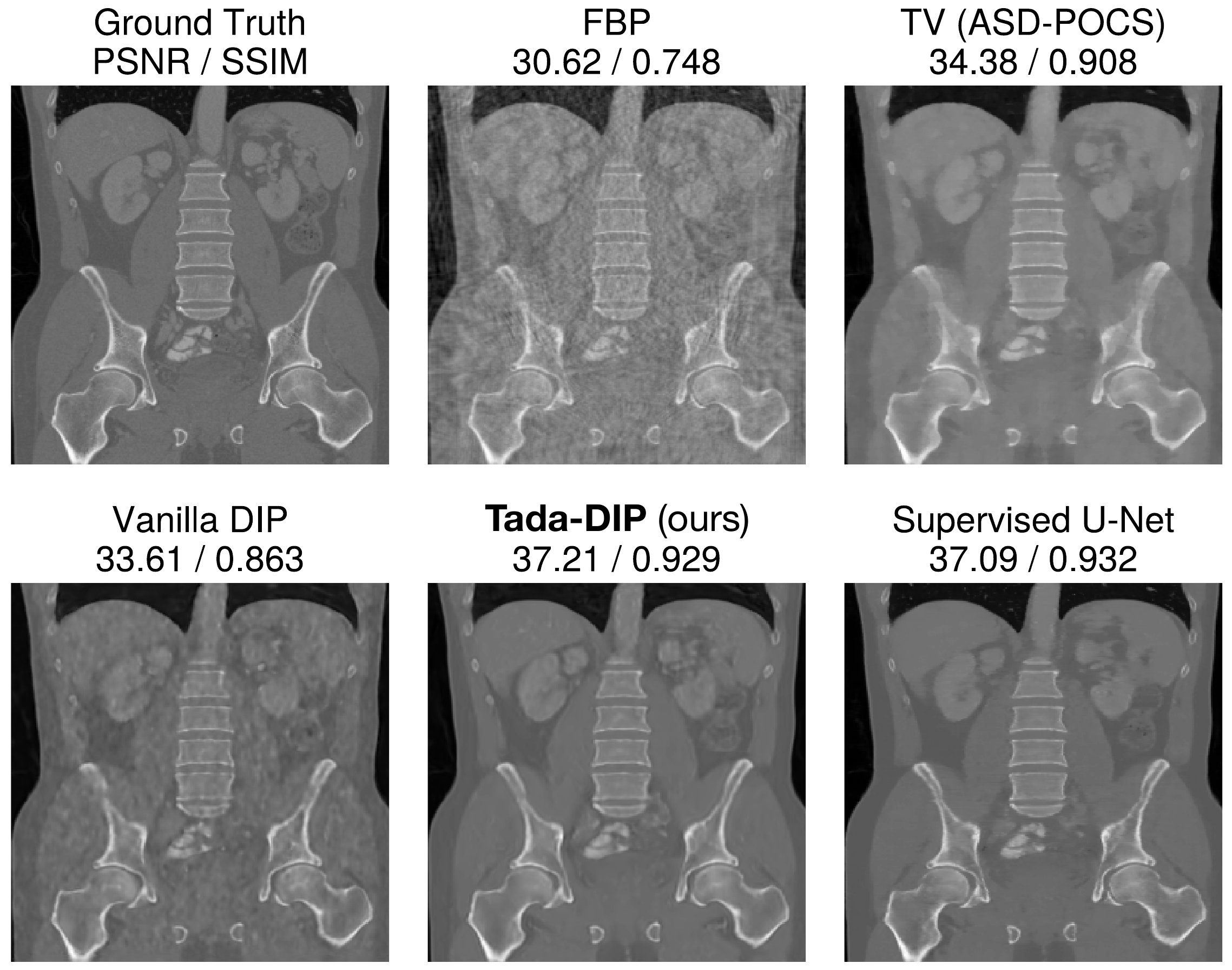}
    \caption{Visualization of coronal slices for 30-view parallel beam reconstruction (same volume as \autoref{fig:qual_comp_30_axial}). Note Tada-DIP's ability to accurately reconstruct fine details, whereas the supervised reconstruction appears overly smooth.}
    \label{fig:qual_comp_30_coronal}
\end{figure}

\begin{figure}[htbp]
    \centering
    \includegraphics[width=\linewidth]{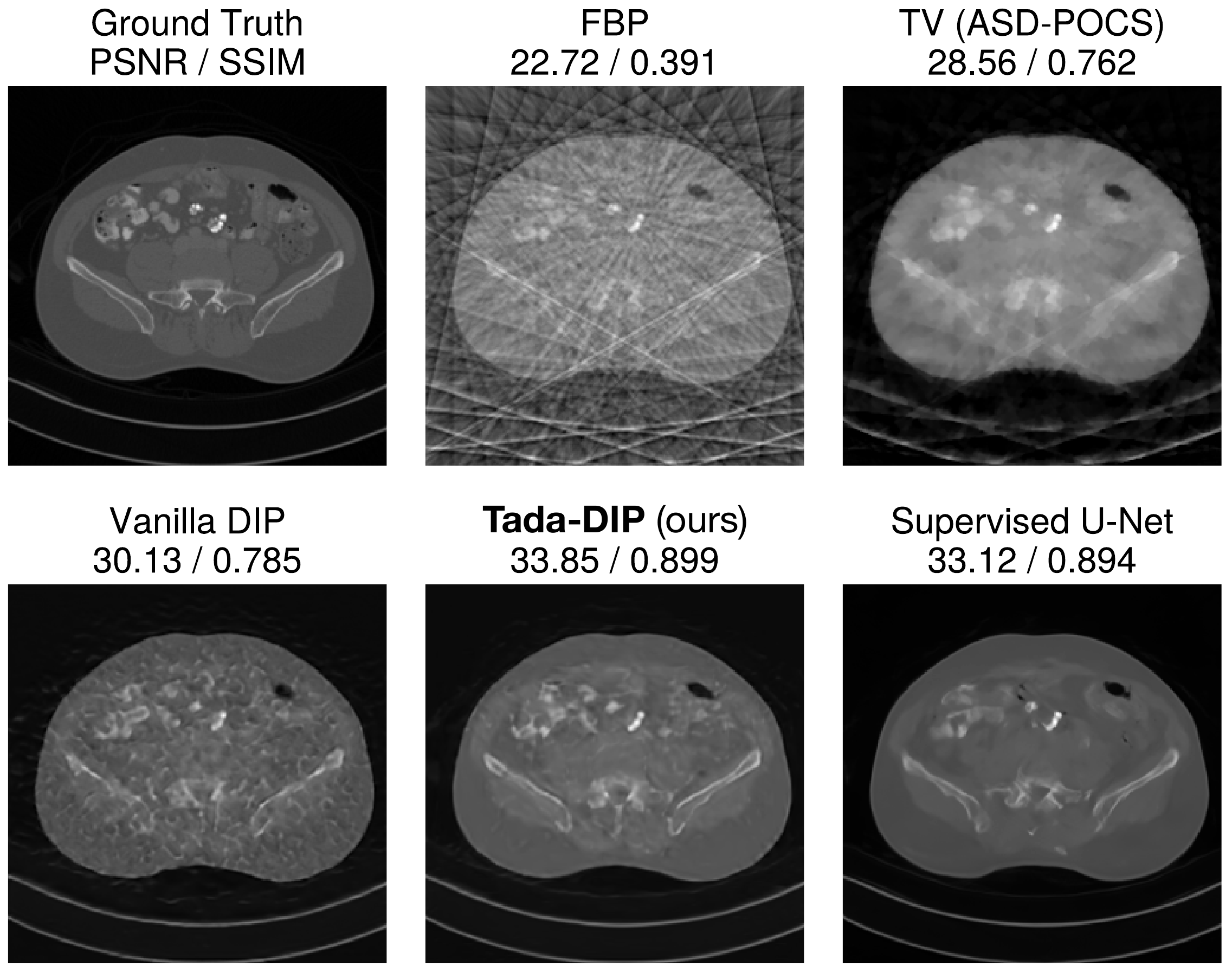}
    \caption{Qualitative comparison of reconstruction methods for 15-view parallel beam CT reconstruction.}
    \label{fig:qual_comp_15_axial}
\end{figure}

\textbf{Qualitative results.\quad}A visualization of the 3D reconstruction produced by Tada-DIP along with the ground truth image for one test case is provided in \autoref{fig:3d_visualization}. Tada-DIP produces a highly faithful reconstruction from just 30 projections, and is able to accurately reconstruct small details, as shown in the accompanying zoom-ins.
Visualizations of slices of the reconstructions produced by Tada-DIP and baseline methods for 30-view reconstruction are provided in \Cref{fig:qual_comp_30_axial,fig:qual_comp_30_coronal}. Visually, Tada-DIP produces very clean reconstructions that are free of the various types of artifacts produced by the other dataless methods. In general, the visual quality of the Tada-DIP reconstruction is similar to that of the supervised reconstruction.
A similar visualization for the 15-view reconstruction problem is shown in \autoref{fig:qual_comp_15_axial}. Even in this extremely challenging setting, the Tada-DIP reconstruction is relatively artifact-free, while the reconstructions from FBP, TV, and Vanilla DIP all exhibit severe artifacts. Again, we find that that the qualitative performance of Tada-DIP is highly similar to that of the supervised network.

\begin{figure}[htbp]
    \centering
    \includegraphics[width=0.7\linewidth]{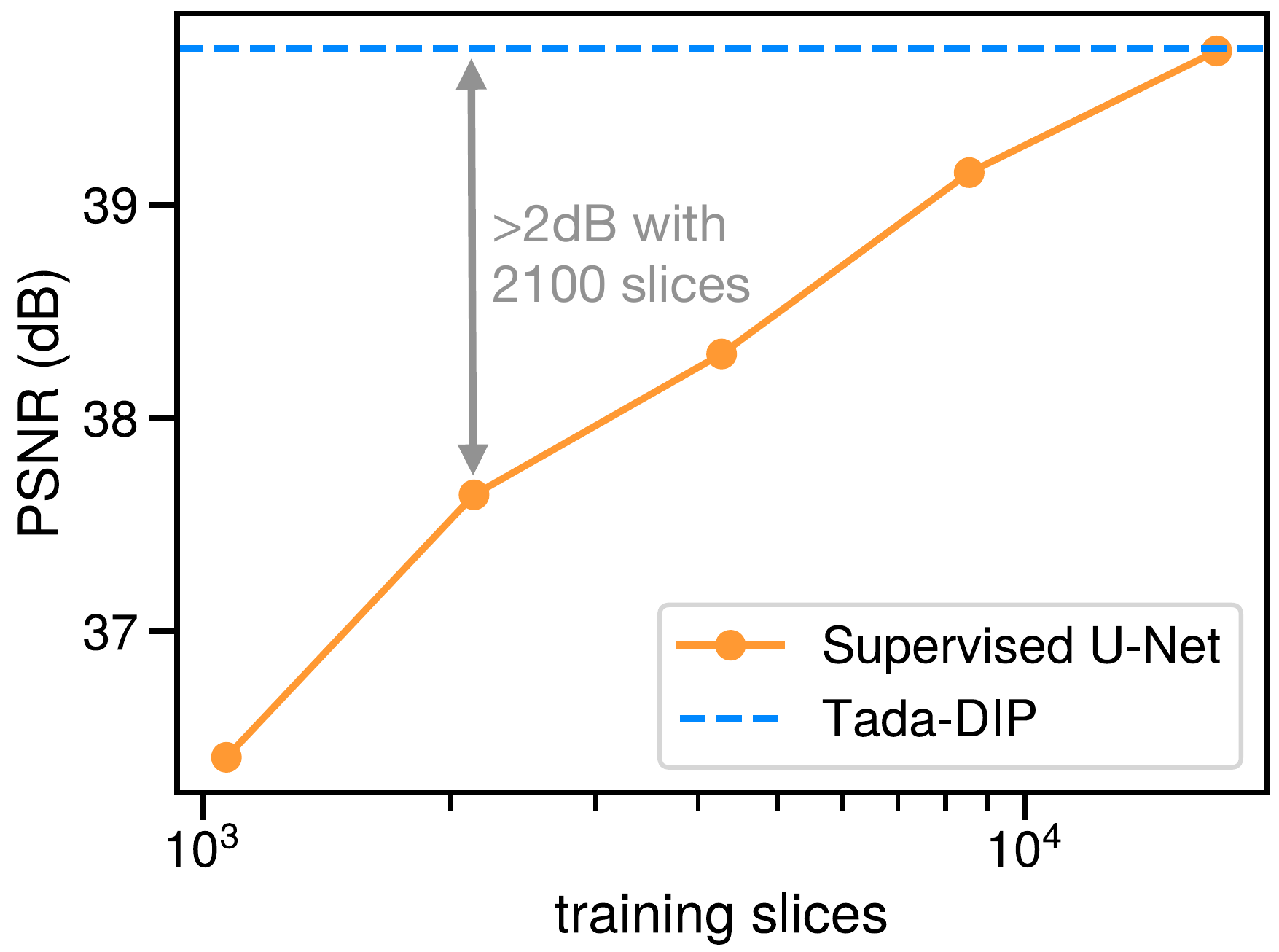}
    \caption{Test performance of a supervised U-Net trained with various dataset sizes compared with our training-data-free algorithm, Tada-DIP.}
    \label{fig:scaling}
\end{figure}

\textbf{Quantitative results.\quad}Tada-DIP's strong performance is also supported by quantitative evaluations. A quantitative comparison of the methods (in terms of PSNR and SSIM) is provided in \autoref{tab:quant_comp}. We find that Tada-DIP substantially outperforms all three unsupervised baselines (by about 2-3 dB across both settings). In fact, the performance of Tada-DIP is essentially identical to that of the supervised U-Net, even for the very difficult 15-view reconstruction problem.
\begin{table}[htbp]
    % \normalsize
    \centering
    \begin{tabular}{lcccc}
    \toprule
    \multirow{2}{*}{\vspace{-0.15cm}Method} & \multicolumn{2}{c}{30 views} & \multicolumn{2}{c}{15 views} \\ \cmidrule(lr){2-3} \cmidrule(lr){4-5}
    & PSNR & SSIM & PSNR & SSIM \\\midrule
    FBP & 29.09 & 0.623 & 24.40 & 0.446 \\
    TV & 35.02 & 0.906 & 30.58 & 0.809 \\
    Vanilla DIP & 37.74 & 0.923 & 32.40 & 0.815 \\
    \midrule
    Supervised U-Net & \underline{39.72} & \textbf{0.955} & \textbf{35.66} & \textbf{0.918} \\
    \midrule
    Tada-DIP (ours) & \textbf{39.73} & \textbf{0.955} & \underline{35.63} & \underline{0.906} \\
    \bottomrule \\
    \end{tabular}
    \caption{Quantitative comparison of Tada-DIP and baseline methods for sparse-view 3D CT reconstruction in terms of PSNR (in dB) and SSIM (with values in $[0,1]$).}
    \label{tab:quant_comp}
    \vspace{-0.75cm}
\end{table}
We acknowledge that this result may be surprising, since the supervised U-Net was trained on a large dataset, whereas Tada-DIP is a dataless method. However, we note that a number of previous studies have reported that untrained networks can match the performance of data-driven methods across diverse tasks, including image reconstruction~\cite{zs-ssl,liang2025analysis}, deblurring~\cite{BID}, and denoising~\cite{alkhouri2024image}. Of course, with sufficient training data, supervised learning may be expected to outperform zero-shot methods. To demonstrate this, we compare Tada-DIP against supervised networks trained with progressively larger subsets of the entire dataset. The test performance of the supervised networks is plotted against the dataset size in \autoref{fig:scaling}. For the supervised method, the performance (in terms of PSNR) follows an approximate power law in the size of the dataset, and only matches Tada-DIP when the entire dataset is used. 
Based on this study, we can infer that supervised learning should outperform Tada-DIP with sufficient data. However, in practice, collecting the required amount of data may be prohibitively expensive. Additionally, the performance of supervised learning can suffer substantially when there is a mismatch between the training and testing data~\cite{heckel2024robust}, while DIP-based methods should be much more robust. 
\begin{figure}[htbp]
    \centering
    \includegraphics[width=0.7\linewidth]{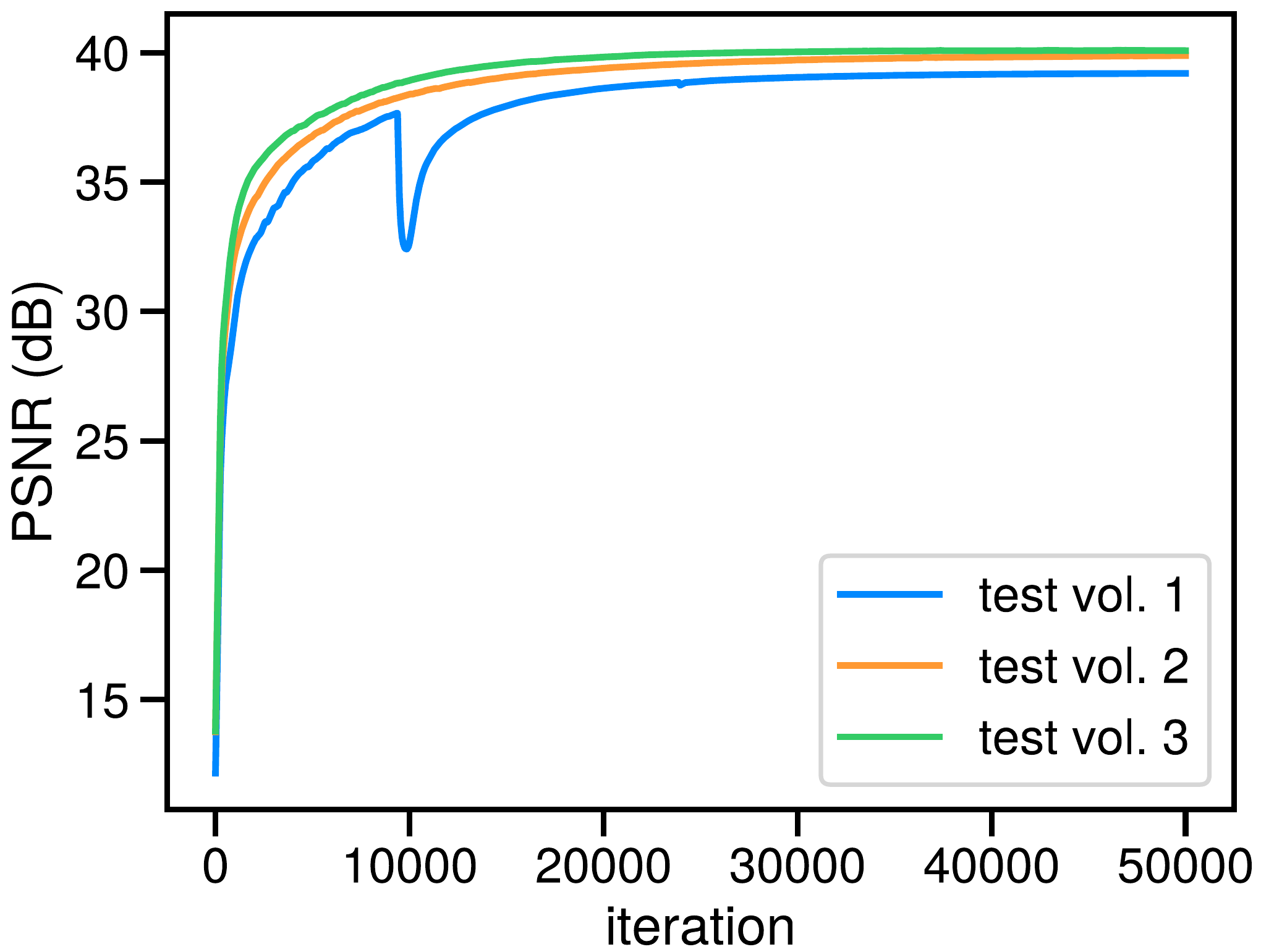}
    \caption{Tada-DIP reconstruction performance vs. number of optimization iterations for three test volumes.}
    \label{fig:convergence}
\end{figure}

Finally, we demonstrate the robustness and convergence of Tada-DIP. The performance of Tada-DIP (in terms of PSNR) versus the number of optimization iterations for the three test volumes is shown in \autoref{fig:convergence}. The PSNR curves show that Tada-DIP generally enjoys smooth convergence and avoids overfitting even after 50000 optimization iterations.

\section{Conclusion}

In this study, we introduced Tada-DIP, a fully 3D Deep Image Prior method that produces high-quality 3D image reconstructions from undersampled measurements without external training data. We empirically validated Tada-DIP by reconstructing 3D X-ray CT images (with resolution 256$^3$) from very sparse measurements (30 and 15 views). Tada-DIP substantially outperformed all dataless baselines, producing reconstructions on par with those of a network trained with supervision and a significant amount of fully-sampled training data.

These results demonstrate that Tada-DIP is a promising method for 3D image reconstruction. This finding also introduces many potential avenues for future research. The first clear direction is performing theoretical analysis of the proposed scheme. While our empirical investigations show that Tada-DIP generally enjoys stable convergence, we do not currently offer a theoretical guarantee. Ideally, such an analysis would also reveal why Tada-DIP avoids overfitting (i.e. converges to a better solution than Vanilla DIP).

Another clear direction is to validate the proposed approach on additional imaging tasks and modalities. While we only investigated 3D X-ray CT in this study, we believe that Tada-DIP would also be effective for other tasks such as accelerated MRI reconstruction. Finally, studying the scalability of Tada-DIP to even larger image reconstruction problems is an important future direction. While the $256^3$ volumes reconstructed in the present study are among the largest reconstructions performed with DIP, applying the proposed method to even higher-resolution images would help to demonstrate its broad practical applicability.

\bibliographystyle{IEEEtran}
\bibliography{ref}

@article{heckel2024robust,
  title={Deep learning for accelerated and robust {MRI} reconstruction},
  author={Heckel, Reinhard and Jacob, Mathews and Chaudhari, Akshay and Perlman, Or and Shimron, Efrat},
  journal={Magnetic Resonance Materials in Physics, Biology and Medicine},
  volume={37},
  number={3},
  pages={335--368},
  year={2024},
  publisher={Springer}
}

@ARTICLE{saisurvey,
  author={Ravishankar, Saiprasad and Ye, Jong Chul and Fessler, Jeffrey A.},
  journal={Proceedings of the IEEE}, 
  title={Image Reconstruction: From Sparsity to Data-Adaptive Methods and Machine Learning}, 
  year={2020},
  volume={108},
  number={1},
  pages={86-109},
  doi={10.1109/JPROC.2019.2936204}}

@article{ctsurvey,
  title={Multi--detector row {CT} systems and image-reconstruction techniques},
  author={Flohr, Thomas G and Schaller, Stefan and Stierstorfer, Karl and Bruder, Herbert and Ohnesorge, Bernd M and Schoepf, U Joseph},
  journal={Radiology},
  volume={235},
  number={3},
  pages={756--773},
  year={2005},
  publisher={Radiological Society of North America}
}

@ARTICLE{fesslermri,
  author={Fessler, Jeffrey A.},
  journal={IEEE Signal Processing Magazine}, 
  title={Model-Based Image Reconstruction for {MRI}}, 
  year={2010},
  volume={27},
  number={4},
  pages={81-89},
  doi={10.1109/MSP.2010.936726}}

@inproceedings{li2023deep,
  title={Deep random projector: Accelerated deep image prior},
  author={Li, Taihui and Wang, Hengkang and Zhuang, Zhong and Sun, Ju},
  booktitle={Proceedings of the IEEE/CVF Conference on Computer Vision and Pattern Recognition},
  pages={18176--18185},
  year={2023}
}

@article{zhao2020reference,
  title={Reference-driven compressed sensing {MR} image reconstruction using deep convolutional neural networks without pre-training},
  author={Zhao, Di and Zhao, Feng and Gan, Yongjin},
  journal={Sensors},
  volume={20},
  number={1},
  pages={308},
  year={2020},
  publisher={MDPI}
}

@inproceedings{alkhourisitcom,
  title={{SITCOM}: Step-wise Triple-Consistent Diffusion Sampling For Inverse Problems},
  author={Alkhouri, Ismail and Liang, Shijun and Huang, Cheng-Han and Dai, Jimmy and Qu, Qing and Ravishankar, Saiprasad and Wang, Rongrong},
 year = {2025},
  booktitle={Forty-second International Conference on Machine Learning}
}

@article{alkhouri2025understanding,
  title={Understanding untrained deep models for inverse problems: Algorithms and theory},
  author={Alkhouri, Ismail and Bell, Evan and Ghosh, Avrajit and Liang, Shijun and Wang, Rongrong and Ravishankar, Saiprasad},
  journal={arXiv preprint arXiv:2502.18612},
  year={2025}
}

@inproceedings{mataev2019deepred,
  title={{DeepRED}: Deep image prior powered by {RED}},
  author={Mataev, Gary and Milanfar, Peyman and Elad, Michael},
  booktitle={Proceedings of the IEEE/CVF International Conference on Computer Vision Workshops},
  pages={1--10},
  year={2019}
}

@inproceedings{cheng2019bayesian,
  title={A {B}ayesian perspective on the deep image prior},
  author={Cheng, Zezhou and Gadelha, Matheus and Maji, Subhransu and Sheldon, Daniel},
  booktitle={Proceedings of the IEEE/CVF Conference on Computer Vision and Pattern Recognition},
  pages={5443--5451},
  year={2019}
}

@inproceedings{unet,
  title={{U-Net}: Convolutional Networks for Biomedical Image Segmentation},
  author={Ronneberger, Olaf and Fischer, Philipp and Brox, Thomas},
  booktitle={International Conference on Medical Image Computing and Computer-Assisted Intervention},
  pages={234--241},
  year={2015},
  organization={Springer}
}

@article{gong2018pet,
  title={{PET} image reconstruction using deep image prior},
  author={Gong, Kuang and Catana, Ciprian and Qi, Jinyi and Li, Quanzheng},
  journal={IEEE Transactions on Medical Imaging},
  volume={38},
  number={7},
  pages={1655--1665},
  year={2018},
  publisher={IEEE}
}

@article{hashimoto2023fully,
  title={Fully {3D} implementation of the end-to-end deep image prior-based {PET} image reconstruction using block iterative algorithm},
  author={Hashimoto, Fumio and Onishi, Yuya and Ote, Kibo and Tashima, Hideaki and Yamaya, Taiga},
  journal={Physics in Medicine \& Biology},
  volume={68},
  number={15},
  pages={155009},
  year={2023},
  publisher={IOP Publishing}
}

@inproceedings{mayo2024stodip,
  title={{StoDIP}: Efficient {3D} {MRF} image reconstruction with deep image priors and stochastic iterations},
  author={Mayo, Perla and Cencini, Matteo and Pirkl, Carolin M and Menzel, Marion I and Tosetti, Michela and Menze, Bjoern H and Golbabaee, Mohammad},
  booktitle={International Workshop on Machine Learning in Medical Imaging},
  pages={128--137},
  year={2024},
  organization={Springer}
}

@inproceedings{hisham2025family,
  title={Family of Deep Image Prior Networks for Accelerated {3D} {LGE-MRI} Acquisition with Enhanced Reconstruction},
  author={Hisham, Md Hasibul Husain and Elhabian, Shireen and Adluru, Ganesh and Arai, Andrew and Kholmovski, Eugene and Ranjan, Ravi and Dibella, Edward},
  booktitle={Medical Imaging with Deep Learning},
  year={2025}
}

@inproceedings{liu2019image,
  title={Image restoration using total variation regularized deep image prior},
  author={Liu, Jiaming and Sun, Yu and Xu, Xiaojian and Kamilov, Ulugbek S},
  booktitle={ICASSP 2019-2019 IEEE International Conference on Acoustics, Speech and Signal Processing (ICASSP)},
  pages={7715--7719},
  year={2019},
  organization={IEEE}
}

@inproceedings{ulyanov2018deep,
  title={Deep image prior},
  author={Ulyanov, Dmitry and Vedaldi, Andrea and Lempitsky, Victor},
  booktitle={Proceedings of the IEEE Conference on Computer Vision and Pattern Recognition},
  pages={9446--9454},
  year={2018}
}

@inproceedings{chung2023diffusion,
  title={Diffusion Posterior Sampling for General Noisy Inverse Problems},
  author={Chung, Hyungjin and Kim, Jeongsol and Mccann, Michael T and Klasky, Marc L and Ye, Jong Chul},
  booktitle={The Eleventh International Conference on Learning Representations, ICLR},
  year={2023},
}

@article{jin2017deep,
  title={Deep convolutional neural network for inverse problems in imaging},
  author={Jin, Kyong Hwan and McCann, Michael T and Froustey, Emmanuel and Unser, Michael},
  journal={IEEE Transactions on Image Processing},
  volume={26},
  number={9},
  pages={4509--4522},
  year={2017},
  publisher={IEEE}
}

@article{hammernik2018learning,
  title={Learning a variational network for reconstruction of accelerated {MRI} data},
  author={Hammernik, Kerstin and Klatzer, Teresa and Kobler, Erich and Recht, Michael P and Sodickson, Daniel K and Pock, Thomas and Knoll, Florian},
  journal={Magnetic Resonance in Medicine},
  volume={79},
  number={6},
  pages={3055--3071},
  year={2018},
  publisher={Wiley Online Library}
}

@article{lustig2007sparse,
  title={Sparse {MRI}: The application of compressed sensing for rapid {MR} imaging},
  author={Lustig, Michael and Donoho, David and Pauly, John M},
  journal={Magnetic Resonance in Medicine: An Official Journal of the International Society for Magnetic Resonance in Medicine},
  volume={58},
  number={6},
  pages={1182--1195},
  year={2007},
  publisher={Wiley Online Library}
}

@article{alkhouri2024image,
  title={Image reconstruction via autoencoding sequential deep image prior},
  author={Alkhouri, Ismail and Liang, Shijun and Bell, Evan and Qu, Qing and Wang, Rongrong and Ravishankar, Saiprasad},
  journal={Advances in Neural Information Processing Systems},
  volume={37},
  pages={18988--19012},
  year={2024}
}

@article{liang2025analysis,
  title={Analysis of deep image prior and exploiting self-guidance for image reconstruction},
  author={Liang, Shijun and Bell, Evan and Qu, Qing and Wang, Rongrong and Ravishankar, Saiprasad},
  journal={IEEE Transactions on Computational Imaging},
  year={2025},
  publisher={IEEE}
}

@inproceedings{bell2023robust,
  title={Robust self-guided deep image prior},
  author={Bell, Evan and Liang, Shijun and Qu, Qing and Ravishankar, Saiprasad},
  booktitle={ICASSP 2023-2023 IEEE International Conference on Acoustics, Speech and Signal Processing (ICASSP)},
  pages={1--5},
  year={2023},
  organization={IEEE}
}

@article{BID,
  title={Blind image deblurring with unknown kernel size and substantial noise},
  author={Zhuang, Zhong and Li, Taihui and Wang, Hengkang and Sun, Ju},
  journal={International Journal of Computer Vision},
  volume={132},
  number={2},
  pages={319--348},
  year={2024},
  publisher={Springer}
}

@inproceedings{zs-ssl,
  title={Zero-Shot Self-Supervised Learning for {MRI} Reconstruction},
  author={Yaman, Burhaneddin},
  booktitle={International Conference on Learning Representations},
  year={2022}
}

@article{mayo_dataset,
  title={Low-dose {CT} image and projection dataset},
  author={Moen, Taylor R and Chen, Baiyu and Holmes III, David R and Duan, Xinhui and Yu, Zhicong and Yu, Lifeng and Leng, Shuai and Fletcher, Joel G and McCollough, Cynthia H},
  journal={Medical Physics},
  volume={48},
  number={2},
  pages={902--911},
  year={2021},
  publisher={Wiley Online Library}
}

@article{aapm_dataset,
  title={Low-dose {CT} for the detection and classification of metastatic liver lesions: results of the 2016 low dose {CT} grand challenge},
  author={McCollough, Cynthia H and Bartley, Adam C and Carter, Rickey E and Chen, Baiyu and Drees, Tammy A and Edwards, Phillip and Holmes III, David R and Huang, Alice E and Khan, Farhana and Leng, Shuai and others},
  journal={Medical Physics},
  volume={44},
  number={10},
  pages={e339--e352},
  year={2017},
  publisher={Wiley Online Library}
}

@article{leap,
  title={Differentiable forward projector for {X}-ray computed tomography},
  author={Kim, Hyojin and Champley, Kyle},
  journal={arXiv preprint arXiv:2307.05801},
  year={2023}
}

@article{asd-pocs,
  title={Image reconstruction in circular cone-beam computed tomography by constrained, total-variation minimization},
  author={Sidky, Emil Y and Pan, Xiaochuan},
  journal={Physics in Medicine \& Biology},
  volume={53},
  number={17},
  pages={4777},
  year={2008},
  publisher={IOP Publishing}
}

\end{document}